# Design of defect spins in piezoelectric aluminum nitride for solid-state hybrid quantum technologies

Hosung Seo[1*], Marco Govoni[1,2], and Giulia Galli[1,2]

[1]The Institute for Molecular Engineering, The University of Chicago, Chicago, IL, USA

[2]Materials Science Division, Argonne National Laboratory, Argonne, IL, USA

**Abstract**

Spin defects in wide-band gap semiconductors are promising systems for the realization of quantum bits, or qubits, in solid-state environments. To date, defect qubits have only been realized in materials with strong covalent bonds. Here, we introduce a strain-driven scheme to rationally design defect spins in functional ionic crystals, which may operate as potential qubits. In particular, using a combination of state-of-the-art *ab-initio* calculations based on hybrid density functional and many-body perturbation theory, we predicted that the negatively charged nitrogen vacancy center in piezoelectric aluminum nitride exhibits spin-triplet ground states under realistic uni- and bi-axial strain conditions; such states may be harnessed for the realization of qubits. The strain-driven strategy adopted here can be readily extended to a wide range of point defects in other wide-band gap semiconductors, paving the way to controlling the spin properties of defects in ionic systems for potential spintronic technologies.

---

[*] Email: hseo@uchicago.edu

**Introduction**

The idea of realizing and harnessing coherent quantum bits in scalable solid-state environments has attracted widespread attention in the past decade [1]. One of the milestones in the field has been the coherent manipulation of the *single* nitrogen-vacancy (NV) defect spin in diamond [2,3], opening a new era for using atom-like point defects in crystals as solid-state qubits [4-5]. However, inherent difficulties in growing and controlling the lattice of C diamond pose severe limitations to the use of the NV center for scalable quantum technologies, begging the question of whether analogs to this defect exist or can be engineered in other technologically mature materials [6-10], for examples III-V crystals. Exploration of quantum defect spins in functional ionic crystals such as, e.g. *w*-AlN, would potentially lead to new opportunities in scalable quantum technologies, including the realization of strongly-coupled spin-mechanical resonator hybrid quantum systems [11-13]. Indeed, the piezoelectric properties of *w*-AlN make it an ideal system for measuring and controlling the vibrational motion of the lattice [14] and may offer a variety of control schemes for quantum spins [8,13]. The strong spontaneous polarization in w-AlN combined with advanced hetero-structuring and band-gap engineering techniques could add flexibility in designing device functionalities [15]. Interestingly, *w*-AlN has recently gained significant attention as an opto-mechanical system [16,17], making it attractive as a new host crystal for quantum defect spins.

A key step towards building hybrid quantum systems in ionic crystals, and in particular in aluminum nitride, is the identification of localized spin-triplet states akin to the NV center in diamond [6]. Within a crystalline environment, the spin sublevels of a spin-triplet state may be split even in the absence of an external magnetic field and they can be chosen to function as a quantum bit [3]. For optical addressability [3], the defect center should possess excited spin-triplet states, which can lead to spin-selective decays. These conditions are, however, challenging to realize in ionic crystals for several reasons. In an ionic solid, the energy of defect states derived from dangling bonds is usually close to either the conduction or the valence band edge and, in many cases it may be in strong resonance with that of the bulk band edges [6]. In the case of *w*-AlN, another critical constraint stems from doping: this material is naturally *n*-type, similar to several other nitrides [18-20], and attaining stable *p*-type doping is extremely challenging [21-25], effectively constraining the search for a defect qubit to *n*-type *w*-AlN.

Here, using first-principles calculations [26-36] we proposed a strain-driven strategy to design stable spin defects in ionic crystals; in particular, we identified the strain conditions leading to the stabilization of localized spin-triplet defect states in *w*-AlN. We found that by applying moderate uni- and bi-axial strain to the host lattice in the presence of negatively charged N vacancies, we could stabilize spin-triplet states, which are well-localized within the fundamental gap of *n*-type *w*-AlN. Our calculations also predicted that these defect spins have a number of spin-conserved excited states, which could be used to optically address the spins [3], thus making them strong candidates for solid-state implementations of quantum bits in piezoelectric aluminum nitride.

## Results

**Spin-state and stability of the native defects in *w*-AlN.** We determined the charge state ($q$) of a given defect $D(q)$ as a function of the Fermi level ($E_F$) by computing the defect formation energy ($E_f^{D(q)}$) [37]:

$$E_f^{D(q)}(E_F) = E^{D(q)} - E^H - \sum n_i \mu_i + q(E_F + E_V) + E_{corr}^D(q), \qquad (1)$$

where $E^{D(q)}$ and $E^H$ are the total energies computed for supercells with and without a defect, respectively; $\mu_i$ ($i$=Al, N, O) is the chemical potential and $E_F$ is referred to the valence band edge ($E_V$). The last term on the right hand side of equation (1) corrects for artificial electrostatic interactions present in calculations with finite-size supercells and we used a correction scheme developed by Freysoldt, Neugebauer, and Van de Walle [38,39]. Figure 1(a) reports $E_f^{D(q)}$ as a function of the Fermi energy for the native defects of $w$-AlN as obtained within PBE [26]. The slope of $E_f^{D(q)}(E_F)$ yields the charge state of the defect [37]; a change in slope indicates a change in the charge state of the most stable defect, as a function of the energy of the Fermi level. The spin of each charge state is also indicated in the figure.

We first considered Al-vacancy ($V_{Al}$)-related defects, including complexes with an N vacancy ($V_{Al}V_N$) and an O impurity at the N site ($V_{Al}O_N$). Paramagnetic states of cation-vacancy-related defects were previously investigated in III-nitrides, in particular GaN [40,41]. The origin of the high-spin states of these cation-vacancy-related defects was attributed to the strong electron localization of the N $2p$ states leading to a large exchange splitting [40,41]. Furthermore, two recent theoretical papers suggested that the paramagnetic state of cation vacancy – oxygen impurity complexes in AlN and GaN may potentially operate as quantum bits, similar to the NV center in diamond [42, 43]. Consistent with previous calculations [40-43], we found that the $V_{Al}$-related defects possess a variety of effective electron spins as shown in Fig. 1(a). However, our calculations showed that the paramagnetic ground states associated with $V_{Al}$-related defects may not be suitable for quantum bit applications for several reasons: those defects with spin states higher than 1/2 are stable in $p$-doped AlN, which is extremely challenging to achieve [20,21]; on the other hand, the defects stable in *n*-type AlN have spin zero. Additionally, the dangling bonds left by the presence of an Al vacancy have mainly N 2*p* character and their position in energy is very close to $E_V$ [40,41]. In some cases, e.g. for the $V_{Al}O_N$ complex [42,43], the active occupied spin-orbital states are in strong resonance with the valence band due to the large exchange splitting. Therefore, we excluded the $V_{Al}$-related defects of *w*-AlN from further consideration in the present work. However we do not exclude $V_{Al}$-related defects from all possible potential qubit operations. For instance, there may exist excited-state triplets, which may be suitable for quantum operations [44].

We now turn to the discussion of anion defects, i.e. $V_N$ and $O_N$, from which the main conclusion of our work will be drawn. In this case, we also obtained results using a level of theory higher than PBE and, in particular, we adopted the PBE0 hybrid functional [27,28] whose choice is extensively justified in the method section. In addition, we carried out calculations using many-body perturbation theory at the $G_0W_0$ level [33-35]. We first note that $O_N$ has no effective spins in its ground state since substitutional oxygen either donates an electron to the host (*q* = +1) or traps an electron (*q* = -1), leading to a filled DX state [45-47]. On the other hand, we found that $V_N$ has S=1/2 and 3/2 ground states for *q* = 0 and -2, respectively, in the *n*-type region, as shown in Fig. 1(a) (PBE results) and (b) (PBE0 results) [25,27]. These findings indicate that the S=3/2 state of $V_N^{2-}$ might be of potential interest as a defect qubit, but it might be difficult to realize, as highly *n*-doped *w*-AlN is difficult to obtain [47,48]. Interestingly, the $V_N^0$ S=1/2 state has been recently detected using electron paramagnetic resonance [18,19]. Based on these results, we considered the Fermi level range in the vicinity of the stability region of $V_N^0$, which might be accessible in experiments. Interestingly, we found that there is a meta-stable state of $V_N^-$ with S=1.

In Fig. 1 we show the *defect-molecule* model [49,50] for $V_N^-$, with the four active Al $sp^3$ dangling bonds, $\{\varphi_i, i=1 \text{ to } 4\}$. The symmetry-adapted molecular orbitals $\{a_1(1), a_1(2), e_x, e_y\}$[51] present in an ideal $C_{3v}$ environment are shown in Fig. 1(c). In such an environment, the ground state configuration of $V_N^-$ is $a_1(1)^2 a_1(2)^2$, which is a spin-singlet. Additionally, attaining the $a_1(1)^2 a_1(2)^1 e^1$ spin-triplet state is possible due to the Hund's rule coupling. Interestingly, both the S=0 and S=1 configurations undergo strong Jahn-Teller (JT) [52] distortions (see Fig. 1(c)). A tight-binding model [53] with hopping matrix elements, -$t_{ij}$, between states $\varphi_i$ and $\varphi_j$ provides a simple picture, and for the S=0 configuration it is easy to show that the $a'(2)$ state is pushed down in energy due to the hybridization with the $a'(3)$ state, thus giving rise to an energy gain. On the other hand, the JT distortion for the S=1 state results in a different orbital ordering as $\Delta t_\parallel \equiv t_{23} - t_{34} > 0$, and the $a'(3)$ state is pushed above $a''$, diminishing the hybridization between $a'(3)$ and $a'(2)$. Our calculations also showed that $a''$ is lowered in energy, thus reducing the energy gap between $a'(2)$ and $a''$ and the S=1 state is stabilized according to the Hund's rule.

**Strain-driven design of defect spins.** We found that in the absence of any strain the S=1 state is higher in energy than the S=0 one by only 55 meV and 7 meV, at the PBE and PBE0 levels of theory, respectively. Note that 7 meV is within the range of our numerical errors, estimated to be ~ 20 meV/defect at most (see supplementary information). This means that the two spin states (S=0 and S=1) of $V_N^-$ are approximately degenerate in energy within PBE0, although they are associated with two distinct JT-distorted structures, as shown in Fig. 1(c). These results suggest that one may engineer the relative stability of the two spin states by straining the lattice. Strain effects have been extensively explored in nitrides and strain levels up to 4% are considered realistic [54,55]. Fig. 2(a) reports the energy difference between the S=1 and S=0 states of $V_N^-$ as a function of compressive uniaxial strain *applied to the host AlN lattice* along the $[11\bar{2}0]$ direction, and shows that the S=1 state is significantly lower in energy than the S=0 state. Even at a small compressive strain of -3%, the S=1 state is lower in energy by about 0.25 eV than the S=0 state within PBE0. This energy difference is well outside our numerical errors (see Supplementary Information) and we expect the energy difference to be representative of the free energy difference as well, since vibrational contributions to the free energy will be almost identical in the two spin configurations (see Supplementary Information). The uniaxial strain effect is twofold. The strained lattice environment is favorable for the S=1 state as the lattice distortion is compatible with the $a_1+e$ JT mode shown in Fig. 1(c). Furthermore the host lattice expands in the [0001] and $[1\bar{1}00]$ directions according to the Poisson's ratio effect [56], which is unfavorable for the S=0 state because it tends to pull apart the $Al_3$ and $Al_4$ atoms, as well as the $Al_1$ and $Al_2$ atoms. Fig. 2(b) reports the defect level diagram of the $V_N^-$ S=1 state under -1% uniaxial strain calculated at the $G_0W_0$@PBE level, showing the occupied $a'(2)$ and $a''$ levels being located almost 2 eV below the conduction band minimum (CBM). We also note that in all cases $a'(2)$ and $a''$ are almost degenerate in energy. As shown in Table 1, the band gap of AlN increases by 0.18 eV within $G_0W_0$@PBE, as the uniaxial strain changes from zero to -3%. However, the energy location of the occupied levels $a'(2)$ and $a''$ does not vary and it remains 3.9 eV above $E_V$ ($v$ in Table 1), indicating that the $^3A''$ spin-triplet state becomes more localized as the compressive uniaxial strain is increased, a beneficial effect for potential quantum information applications.

We further explored the possible stability of the S=1 $V_N^-$ defect by considering biaxial strain. Fig. 3(a) reports the energy difference between the S=1 and S=0 states of $V_N^-$ as a function of biaxial strain applied in the (0001) plane, showing the stabilization of the S=1 state above 3% strain, within the PBE0 approximation. Under 4% biaxial strain, the S=1 state is lower in energy by 80 meV than the S=0 state in PBE0, a result again outside the numerical errors and possible temperature effects (see Supplementary Information). We note that under the biaxial strain considered in Fig. 3(a), the lowest-energy geometry for the S=0 state is again the $C_{1h}^{(e)}$ JT-distorted structure, with the same orbital ordering shown in Fig. 1(c). However, an increase of biaxial strain to 2% induces a structural transition in the S=1 state of $V_N^-$, and the defect symmetry changes from $C_{1h}$ to $C_{3v}$. Moreover, the structural transition is accompanied by an $a_1$-type displacement, in which Al$_4$ (see Fig. 1(b)) is moved out of the (0001) plane and up in the z-direction, for example by 0.3 Å at 3% biaxial strain. Let us consider a tight-binding model in the $C_{3v}$ symmetry. The orbital energy of the e-states and the $a_1(2)$ state are $t_\parallel$ and $-t_\parallel + \sqrt{t_\parallel^2 + 3t_\perp^2}$, respectively; here, $t_\parallel$ (= $t_{23}$ = $t_{34}$ = $t_{24}$) and $t_\perp$ (= $t_{12}$ = $t_{13}$ = $t_{14}$) are the in-plane and out-of-plane hopping constants, which are decreased and increased, respectively, under tensile biaxial strain. Therefore, application of such strain could reverse the orbital order between $e_{x,y}$ and $a_1(2)$ in $V_N^-$ beyond a certain critical strain level and lead to the formation of a $^3A_2$ S=1 spin-triplet state, according to Hund's rule. Figure 3(b) reports the $G_0W_0$ quasi-particle electronic structure of the S=1 state of $V_N^-$ under 3% biaxial strain, showing the doubly degenerate e-states with two spin-up electrons localized in the band gap of w-AlN. We found that the band gap of w-AlN is decreased from 5.94 eV to 5.35 and 5.10 eV under biaxial strain of 3% and 4%, respectively, within the $G_0W_0$@PBE approximation (see Table 1). However, in both cases, the e-states are well-localized and located deep in the gap, 1.8 eV (3% strain) and 1.7 eV (4% strain) below the CBM, within $G_0W_0$@PBE.

**Hyperfine coupling between the $V_N^-$ defect spin and Al nuclear spins.** Electron paramagnetic resonance (EPR) [57] is a powerful experimental technique to detect and identify paramagnetic impurities in solids. EPR measurements yield hyperfine parameters, which are mainly determined by the interaction between an impurity's electron spin and the surrounding nuclear spins, thus these measurements play a key role in the identification of point-like paramagnetic impurities in solids [57]. Son and co-workers resolved the detailed hyperfine structure of an electron spin-1/2 in *w*-AlN, mostly interacting with four $^{27}$Al nuclei (nuclear spin I=5/2, 100% natural abundance), and they identified its origin to be $V_N^0$ with the help of *ab-initio* density functional calculations [18]. In Table 2, we report the computed principal values of the hyperfine tensor of $V_N^0$ and we compare them with the previous theoretical and experimental data. Our results are in good agreement with all the previous results, supporting the interpretation of the resolved hyperfine structure being derived from $V_N^0$.

Similar to $V_N^0$ in *w*-AlN, the $V_N^{-1}$ electron spin S=1 may also interact with the nearest four Al atoms ($Al_{1-4}$ shown in Fig. 1). We calculated the principal values of the hyperfine tensor of $V_N^-$ (S=1) as a function of strain and the results are reported in Fig. 4. Interestingly, the hyperfine parameters exhibit high sensitivity to the applied strain. For the uniaxial case, the hyperfine parameters ($A_{xx}$, $A_{yy}$, and $A_{zz}$) for $Al_2$ decrease by -75 MHz as the compressive uniaxial strain is increased from 0 to -3%, while those of $Al_{1,3,4}$ increase. To understand this trend, we show in Fig. 4 the isotropic Fermi contact term as a function of strain along with the hyperfine parameters.

We found that the Fermi contact term, which is mainly determined by the electron spin density localized at the corresponding Al site, is mostly responsible for the change of the hyperfine parameters as a function of strain. Fig. 4 (a) to (c) indicate that as a result of uniaxial strain in the $V_N^-$ S=1 state the spin density is transferred from $Al_2$ to the other Al atoms.

Interestingly, the hyperfine parameters as functions of biaxial strain show a more drastic change. The hyperfine parameters for $Al_1$ decrease by more than 250MHz and become negative while those of the other basal Al nuclei ($Al_{2-4}$) significantly increase. This is due to the structural transition of the defect geometry from $C_{1h}$ to $C_{3v}$ beyond the 2% biaxial strain, as previously discussed. As a result of the transition, beyond 2% strain the $Al_2$, $Al_3$, and $Al_4$ nuclei become symmetrically equivalent and hence their hyperfine parameters become equal, as shown in Fig. (e) and (f). The negative hyperfine parameters of $Al_1$ are due to the negative spin density localized near the $Al_1$ atom, which makes the Fermi contact term negative. Before the structural transition occurs, the spin density is distributed almost equally over the four nearest Al nuclei as it is derived from the a'(2) and a" orbitals in the $C_{1h}$ symmetry. The structural transition to $C_{3v}$ induces, however, a change in the orbital ordering and the spin density is derived from the $e_x$ and $e_y$ orbitals, which are mainly localized at the basal plane of $Al_2$, $Al_3$, and $Al_4$, leading to increased Fermi contact terms as shown in Fig. 4 (e) and (f).

**Spin-conserved intra-defect excitations in nitrogen vacancy spins.** One of the most important properties of the NV center in diamond is the single-spin optical addressability, which relies on the presence of an excited $^3E$ spin-triplet state and its spin-selective decay [3]. Similar to the diamond NV center, we found that in *w*-AlN the atomic configuration in which the negatively charged N vacancy has a S=1 ground state also exhibits S=1 excited states; hence these states could be obtained by a spin-conserving optical excitation from an occupied defect orbital to an empty defect orbital, as shown in Fig. 2(b) and Fig. 3(b). Some of the empty defect levels are located slightly above the CBM. We note, however, that the lowest-lying empty defect-orbitals are not in resonance with the CBM; they remain localized, as shown by their dispersion-less character in our computed band structure (not shown) due to the following reasons. The lowest conduction band of $w$-AlN is a single parabolic band centered at $\Gamma$. The next available conduction states appear close the K and L points and they are located at 0.9 and 1.1 eV above the CBM, respectively, according to previous GW calculations [58], and consistent with our GW results. Furthermore we found that the CBM mainly exhibits a nitrogen $p$ character, leading to a negligible hybridization with the aluminum dangling bonds created by the $V_N^-$ defect.

The ZPL was obtained by carrying out calculations of total energy differences ($\Delta$SCF calculations) within the PBE and PBE0 approximation [59,60]. We first describe possible intra-defect spin-conserving excitation in the spin-down channel, which may be similar to the excitation scheme of the NV center in diamond [60]. In the case of uniaxial strain, we promoted an electron from $a'(1)$ to $a'(2)$ and the resulting configuration corresponds to the optically excited $^3A''$ spin-triplet state. The ZPL is rather insensitive to the amount of uniaxial strain, up to 3%, and it is calculated to be ~ 3.2 eV within PBE. A similar excitation for $V_N^-$ under biaxial strain, shown in Fig. 3(b), corresponds to a transition to the excited $^3E$ spin-triplet state; for the ZPL we found 3.35 and 3.25 eV under 3% and 4% biaxial strain, respectively within PBE. We note, however, that these near ultra-violet excitations may lead to photo-ionization of $V_N^-$ to $V_N^0$ considering the corresponding (0/-1) charge transition level is about 1.1 eV below the conduction band edge, as shown in Fig. 1(b). In the spin-up channel, however, the ZPLs from *a″* to *a′*(3) under uniaxial strain, and from *e* to $a_1$(2) under biaxial strain, are expected to be at much lower energies, in the near infrared range. For these cases, we used the PBE0 hybrid functional to better estimate the ZPLs [60]. We find that the spin-up ZPLs are 0.83 eV and 0.89 eV for -1% uniaxial strain and 3% biaxial strain, respectively.

**Discussion**

We proposed a strain-driven defect design scheme to obtain point defects with localized spin-triplet ground states for implementation of spin qubits in piezoelectric aluminum nitride. We found that negatively charged nitrogen vacancies exhibit localized spin-triplet states in *n*-type aluminum nitride under realistic strain conditions. Nitrogen vacancies are naturally incorporated in aluminum nitride during crystal growth and they are known to be the main source of the intrinsic *n*-type behavior of aluminum nitride [18-21], thus making the nitrogen vacancy spins easily amenable to experimental investigations [18,19]. Extra *n*-type doping control might be achievable by introducing substitutional oxygen impurities ($O_N$) during crystal growth [24]. As shown in Fig. 1(a) and (b), $O_N$ can donate electrons in the region where $V_N^0$ is stable and provide the extra charge necessary to form $V_N^-$.

In addition to nitrogen vacancies, $V_{Al}$-related defects are also commonly observed in experiments [22,23]. Our calculations showed that such defects could in principle introduce a variety of electron spins in the host lattice, which would make it difficult to isolate a single defect spin for qubit applications. However, as shown in Fig. 1(a), the $V_{Al}$-related defects are magnetically passivated (i.e. S=0) in the stability region of $V_N^-$, thus allowing for the $V_N^-$ spins to be easily isolated for qubit applications.

We showed that both the S=0 and S=1 configurations of $V_N^-$ undergo static Jahn-Teller (JT) [52] distortions as shown Fig. 1(c) and the relative stability of the S=0 and S=1 states can be controlled by applying a uniaxial or biaxial strain to the host lattice. Regarding possible temperature effects on the relative stability under strain, dynamic Jahn-Teller effect [52] might be an important factor to be considered in addition to the vibrational entropy effect discussed in the Supplementary Material. We also note that potential dynamic Jahn-Teller effects at elevated temperature might be strain-dependent, as we found the defect structure and its lattice environment change significantly in response to external strain perturbations. An investigation of potential dynamic Jahn-Teller effects will be the subject of future studies.

We also showed that excited spin-triplet states are present for negatively charged nitrogen vacancies and these states may play an important role in potential optical manipulations of the vacancy spins. We found that the zero phonon lines for a spin-conserving intra-defect excitation in the spin-down and spin-up channels of the $V_N^-$ S=1 state are in the near ultra-violet and in the near infrared range, respectively. We pointed out that the near ultra-violet excitation could lead to photo-ionization of the $V_N^-$ defect, thus the near infrared excitation may be more suitable for potential optical manipulation of the $V_N^-$ spin.

The excited triplet may couple to the ground state triplet non-radiatively, similar to the NV center in diamond. We note, however, that additional important issues remain to be considered in order to address the potential optical manipulation of the $V_N^-$ spin. The optical manipulation of the NV center is based on the spin-orbit-induced spin-selective decay through dark singlet states [3]. For the $V_N^-$ system studied here there are open-shell singlet states such as $^1A_1$ or $^1E$ states in the $C_{3v}$ symmetry case and $^1A''$ in $C_{1h}$, which may be close in energy to the $^3A_2$ triplet in $C_{3v}$ and the $^3A''$ triplet in $C_{1h}$, respectively. These single states may play a role in the optical manipulation of the $V_N^-$ spin. However, whether these singlet states are positioned in energy between the ground spin triplet and the excited triplet remains to be seen [61], as well as whether the spin-orbit interaction between the excited triplet states and the singlet states is sufficiently strong to couple them. We further note that the $C_{3v}$ symmetry of the NV center in diamond is responsible for specific selection rules for the spin-selective decay [62]. The $V_N^-$ defect in *w*-AlN has the $C_{3v}$ symmetry only under the biaxial strain, while it has $C_{1h}$ symmetry under uniaxial strain; such symmetry difference may be an important factor in establishing the potential optical manipulation of the $V_N^-$ spin.

It is worth mentioning that a number of alternative strain-driven spin manipulation and readout schemes are being actively developed in the literature [63,64], and the $V_N^-$ spins proposed in our study may lead to excellent platforms for the implementation of strain-driven spin control schemes due to the strong piezoelectricity of the host lattice [8].

**Methods**

*Ab-initio* **charged defect calculations.** We carried out density functional theory (DFT) calculations with semi-local (PBE [26]) and hybrid (PBE0 [27,28]) functionals, using plane wave basis sets (with a cutoff energy of 75 Ry), norm-conserving pseudopotentials [29], and the Quantum Espresso code [30]. In the case of *w*-AlN, the self-consistent Hartree-Fock mixing parameter, as derived using the method of Ref. [28], is 24% [28], which justifies the use of the PBE0 hybrid functional, whose mixing parameter is defined to be 25% (see below). We examined the most common native defects, which may be easily accessible in experiment (including Al [22,23] and N vacancies [18,19], O impurities [24], and cation-anion defect complexes [23,25]). To mimic the presence of isolated defects, we employed supercells with 480 and 96-atoms, when using the PBE and PBE0 approximations, respectively, and full geometry optimizations were performed with both functionals. We sampled the Brillouin zone by a 2×2×2 *k*-point mesh. Numerical errors in terms of supercell size, k-point sampling and the plane wave cutoff energy were examined and are summarized in the supplementary information. We considered 7 different types of defects and 6 to 7 different charge states for each of them, in addition to 2 to 3 spin multiplicities for each case. For all defect states, total energy minimizations were started from three different initial geometries (with symmetry $C_{3v}$ and $C_{1h}$) and the state with the lowest energy was selected as the ground state. In total, we explored approximately 300 different defect states.

The use of PBE0 to describe AlN is appropriate, based on our recent work [28], as the average electronic dielectric constant of AlN is ~4.1 [28] and we expect the optimal amount of the Hartree-Fock mixing to be 1/4.1 = 24%, which is almost identical to the 25% mixing parameter entering the definition of the PBE0 functional. The accuracy of the PBE0 functional to describe pristine AlN was checked for several properties, which are summarized in Table 3 and are all in excellent agreement with experiment. In particular, using the PBE0 functional we calculated the electronic and static dielectric constants of AlN with a combined finite E-field and Berry phase method [65,66]. We calculated the electronic dielectric constants to be 4.06 and 4.22 for $\epsilon_{\infty,\parallel}$ and $\epsilon_{\infty,\perp}$, respectively, to be compared with experimental values of 4.13±0.02 for $\epsilon_{\infty,\parallel}$ and 4.27 ± 0.05 for $\epsilon_{\infty,\perp}$ [67]. For the static dielectric constants, we calculated $\epsilon_{0,\parallel}$ and $\epsilon_{0,\perp}$ to be 7.71 and 8.95, respectively, to be compared with experimental values of 9.18 for $\epsilon_{0,\perp}$ [68] and the result of 8.5 obtained from polycrystalline AlN [69]. Note that PBE0 accurately describes the anisotropic nature of the dielectric constants of AlN. The accuracy of the PBE0 functional to describe the native defects in AlN was checked by calculating defect formation energies and charge transition levels of the N vacancy and O impurity as shown in Fig. 1(b). Recently, similar results using another hybrid functional (HSE) were reported in Ref. [25]; the authors were able to successfully explain the experimentally observed optical absorption and emission of the defects and we verified that our PBE0 results are in almost perfect agreement with those of Ref. [25].

The zero-phonon line (ZPL) was obtained by carrying out calculations of total energy differences (ΔSCF calculations) within the PBE and PBE0 approximation [59,60]. We employed 480-atom and 288-atom supercells along with 75 Ry and 65 Ry plane-wave cutoff energies for the PBE and PBE0 calculations, respectively. By using the PBE functional, we checked the numerical error induced by reducing the supercell size from 480 atoms to 288 atoms and the plane-wave cutoff energy from 75Ry to 65Ry to be around 0.05eV.

**First-principles calculations of hyperfine tensors.** We carried out calculations of hyperfine tensors between $V_N$ spins and nuclear spins in *w*-AlN at the PBE level of theory. The calculations were performed in two steps. First, we calculated the ground-state wavefunctions for $V_N$ using the 480-atom supercell as described in the previous section. Then, we used the gauge-including projector-augmented wave method (GIPAW) of Ref. [70] to calculate the hyperfine tensor, which is comprised of the isotropic Fermi contact term and the anisotropic dipolar coupling term. Numerical convergence in terms of the energy cutoff, the k-point sampling, and the supercell size was checked, with a numerical error in the hyperfine tensor less than 10 MHz. To verify the accuracy of the PBE functional, we calculated the hyperfine parameters of the neutral $V_N$ spin (S=1/2) and compared them to previous theoretical and experimental results [18] as shown in Table 2. We found an excellent agreement between our and previous results. In addition, we found that adding core polarization effects [71, 72] improves the agreement between our results and experiment, thus we included core polarization effects throughout all of our calculations.

**Large-scale many-body GW calculations.** In order to check the robustness of our predictions, we carried out calculations of quasi-particle energies within the $G_0W_0$ approximation [31,32], using the Γ point only and 480-atom supercells with defect geometries optimized at the PBE level of theory. GW calculations were performed utilizing a spectral decomposition technique for the dielectric matrix [33,34] and an efficient contour deformation technique for frequency integration [35], as implemented in the WEST code (www.west-code.org), where the evaluation of virtual electronic states is not required. Calculations carried out with the WEST code started from the results obtained with the semi-local PBE functional and perturbative corrections to the Kohn-Sham eigenvalues were obtained. The massively parallel implementation of the GW method in WEST takes advantage of separable expressions for both the Green's function (G) and the screened Coulomb interaction (W). The newly developed technique for large-scale GW calculations allowed us to explore defective AlN systems of unprecedented size, containing

~2000 electrons. For the band gap of *w*-AlN, we obtained 5.94 eV within the $G_0W_0$@PBE approximation using our 480-atom supercell with a point defect, which is in very good agreement with the previous $G_0W_0$@LDA results of 5.8 eV [58] and 6.08 eV [73] obtained for the pristine bulk.

**Acknowledgements**

We thank David Awschalom for suggesting *w*-AlN as a new potential host material for quantum defect spins. We also thank William Koehl, Jonathan Skone, Matthew Goldey, Márton Vörös, and Eun-Gook Moon for helpful discussions. HS is primarily supported by the National Science Foundation through the University of Chicago MRSEC under award number DMR-1420709. MG and GG are supported by DOE grant No. DE-FG02-06ER46262. Part of this work (MG) was done at Argonne National Laboratory, supported under U.S. Department of Energy contract DE-AC02-06CH1135. This research used resources of the National Energy Research Scientific Computing Center (NERSC), a DOE Office of Science User Facility supported by the Office of Science of the U.S. Department of Energy under Contract No. DE-AC02-05CH11231, resources of the Argonne Leadership Computing Facility, which is a DOE Office of Science User Facility supported under Contract DE-AC02-06CH11357, and resources of the University of Chicago Research Computing Center.

**Author Contributions**

H.S. and G.G. designed the research. Most of the calculations were performed by H.S., with contributions from all authors. M.G. implemented the GW many-body perturbation theory in the WEST code. All authors contributed to the analysis and discussion of the data and the writing of the manuscript.

**Competing Financial Interests statement**

The authors declare no competing financial interests.

# Figures and Figure Legends

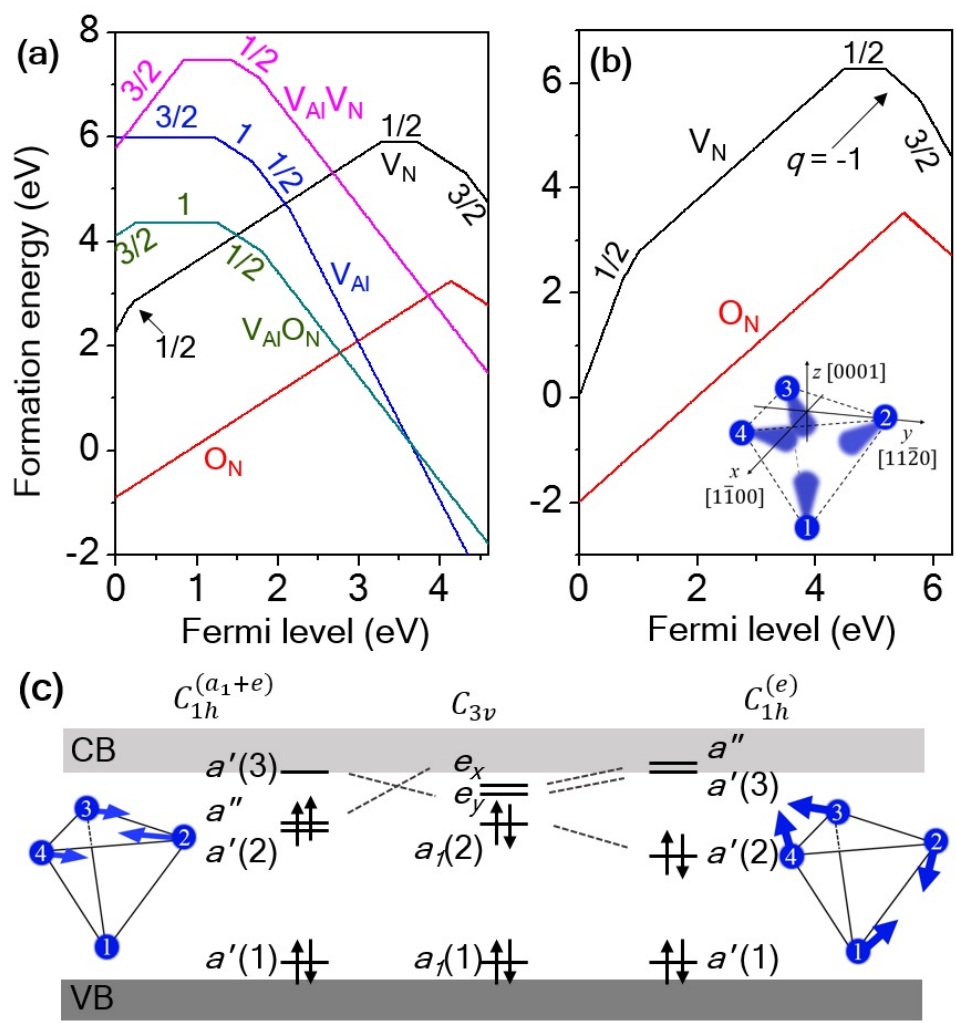

**Figure 1 | Defect formation energy and spin-state of the native defects in *w*-AlN.** Results obtained within PBE (a) and formation energy of $V_N$ and $O_N$ computed within PBE0 (b). The lowest-energy effective spin for a given charge state is also shown. The inset of (b) is a schematic representation of the defect-molecule model of $V_N$. Only the nearest neighbor Al atoms (numbered spheres) around the vacancy site and the Al $sp^3$ dangling bonds are shown for clarity. (c) Symmetry-adopted molecular orbitals and spin configurations of $V_N^-$ under different symmetry environments. The Jahn-Teller distortions for the $S=0$ and the $S=1$ states are described by the $e$ and $a_1+e$ symmetrized displacements, respectively, as schematically shown next to the defect level diagrams.

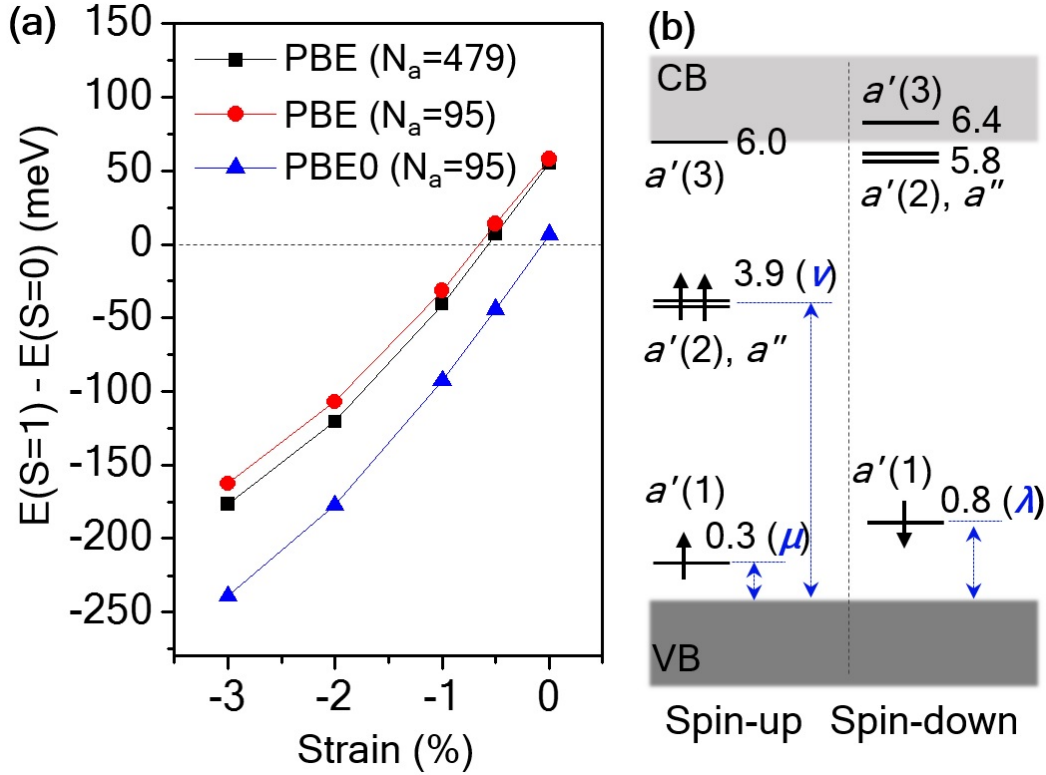

**Figure 2 | Control of the nitrogen vacancy spin using uniaxial lattice strain.** (a) Total energy difference between the S=1 and S=0 states of $V_N^-$ as a function of the compressive uniaxial strain applied along the $[11\bar{2}0]$ direction. (b) The $G_0W_0$@PBE quasi-particle electronic structure of the S=1 state of $V_N^-$ under uniaxial strain of -1%. The number next to a defect level is the location of the state referred to $E_V$. Three constants, $\mu$, $\nu$, and $\lambda$ are used to denote the positions of the occupied defect orbitals (see Table 1).

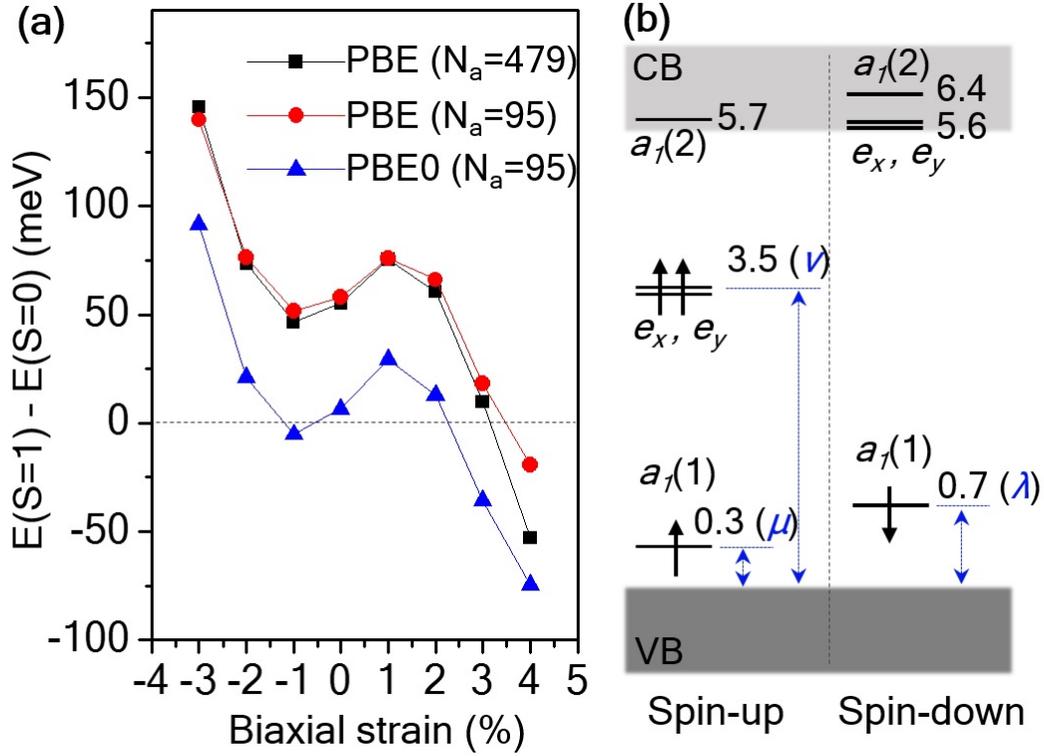

**Figure 3 | Control of the nitrogen vacancy spin using biaxial lattice strain.** (a) Total energy difference between the S=1 and S=0 state of $V_N^-$ as a function of the biaxial strain applied in the (0001) plane. The local defect geometry of the S=0 state maintains a $C_{1h}^{(e)}$ symmetry, as shown in Fig. 1(c), while that of the S=1 state changes from $C_{1h}^{(a1+e)}$ to $C_{3v}$ for biaxial strain larger than 2%. (b) The $G_0W_0$@PBE quasi-particle electronic structure of the S=1 state of $V_N^-$ under 3% biaxial strain. The number next to a defect level is the location of the state referred to $E_V$. Three constants, $\mu$, $\nu$, and $\lambda$ are used to denote the positions of the occupied defect orbitals (see Table 1).

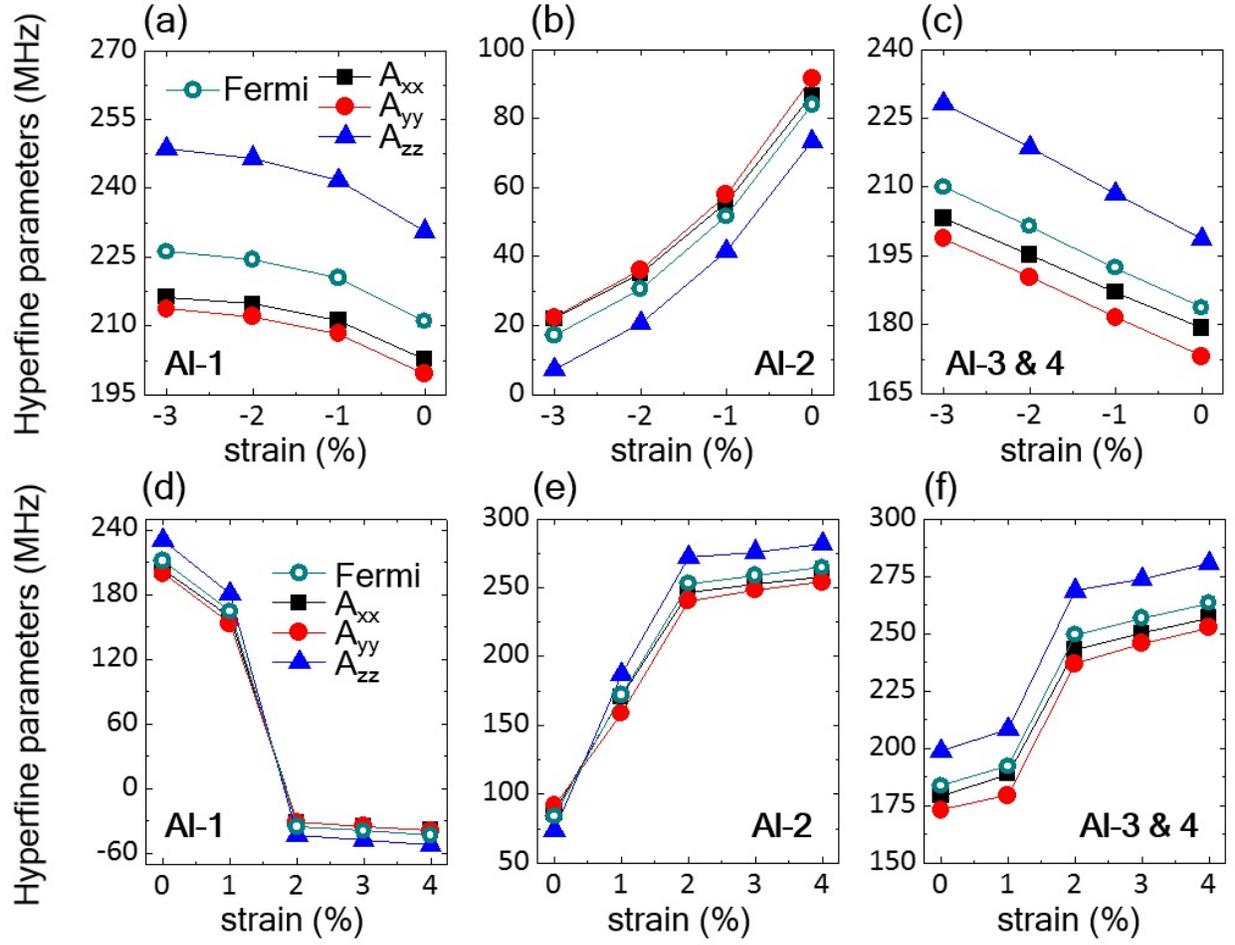

**Figure 4 | Hyperfine parameters of the $V_N^-$ spin.** Principal values (in MHz) of the hyperfine tensors and Fermi contact terms for interaction between the $V_N^-$ electronic spin (S=1) with the nearest four $^{27}$Al nuclei as a function of the uniaxial (a-c) and biaxial (d-f) strains. The four $^{27}$Al nuclei ($Al_1$ to $Al_4$) can be seen in Figure 1. Note that the applied uniaxial strain and the biaxial strain are compressive (negative value) along the [11$\bar{2}$0] direction and tensile (positive value) in the (0001) plane, respectively. The local defect geometry of the S=1 state changes from $C_{1h}^{(a1+e)}$ to $C_{3v}$ for biaxial strain larger than 2%, inducing the drastic change in the hyperfine parameters.

# Table

**Table 1.** Computed electronic properties of the $V_N^-$ S=1 state in *w*-AlN under the uniaxial and biaxial strain discussed in the main text, including the band gap within $G_0W_0$@PBE, the location of the occupied defect orbitals referred to $E_V$ in $G_0W_0$@PBE ($\mu,\nu,\lambda$ shown in Fig. 2(b) and Fig. 3(b)).

| Strain | Band gap (eV) | Defect orbitals (eV) | | |
|---|---|---|---|---|
| | | $\mu$ | $\nu$ | $\lambda$ |
| Strain-free | 5.94 | 0.36 | 3.90 | 0.73 |
| Uniaxial, -1% | 6.01 | 0.34 | 3.91 | 0.77 |
| Uniaxial, -2% | 6.11 | 0.47 | 3.93 | 0.83 |
| Uniaxial, -3% | 6.12 | 0.42 | 3.89 | 0.80 |
| Biaxial, 3% | 5.35 | 0.30 | 3.55 | 0.66 |
| Biaxial, 4% | 5.10 | 0.20 | 3.42 | 0.59 |

**Table 2.** Computed principal values (in mT) of the hyperfine tensors of $V_N^0$ in *w*-AlN at the PBE level of theory, compared to previous theoretical and experimental results [18]. The four nearest $^{27}$Al (I=5/2, 100% natural abundance) of $V_N^0$ shown in Figure 1 are considered. The calculated principal values include core polarization (CP) effects, also reported in the table.

| Center | Atom | | $A_{xx}$ (mT) | $A_{yy}$ (mT) | $A_{zz}$ (mT) | CP (mT) |
|---|---|---|---|---|---|---|
| $V_N^0$ | Al$_1$ | This work | 6.0 | 6.0 | 7.5 | -1.2 |
| | | Theory[18] | 6.4 | 6.5 | 8.0 | |
| | Al$_2$ | This work | 6.4 | 6.4 | 7.7 | -1.1 |
| | | Theory[18] | 6.7 | 6.7 | 8.2 | |
| | Al$_3$ Al$_4$ | This work | 5.3 | 5.3 | 6.9 | -1.2 |
| | | Theory[18] | 5.9 | 6.0 | 7.5-7.6 | |
| EI-1 | Al$_{1-4}$ | Experiment[18] | ~6.0 | ~6.0 | ~7.2 | |

**Table 3.** Computed electronic and structural properties of *w*-AlN using PBE and PBE0 compared to the experimental values, including the lattice constant, the direct band gap ($E_g$), the crystal field splitting ($\Delta_{CF}$), and the electronic and static dielectric constants.

| | Lattice parameters | | | Electronic properties | | Dielectric constants | |
|---|---|---|---|---|---|---|---|
| | *a* (Å) | *c/a* | *u* | $E_g$ (eV) | $\Delta_{CF}$ (meV) | Electronic ($\epsilon_{\infty,\parallel}/\epsilon_{\infty,\perp}$) | Static ($\epsilon_{0,\parallel}/\epsilon_{0,\perp}$) |
| PBE | 3.130 | 1.603 | 0.382 | 4.09 | -201 | 4.42 / 4.63 | 8.33 / 9.77 |
| PBE0 | 3.107 | 1.601 | 0.382 | 6.35 | -223 | 4.06 / 4.22 | 7.71 / 8.95 |
| Experiment | 3.110 [36] | 1.601 [36] | 0.382 [36] | 6.0-6.3 [36] | -230 [36] | 4.13±0.02 [67] / 4.27 ± 0.05 [67] | (not known) / 9.18 [68], 8.5 (poly-crystalline) [69] |